\documentclass[
    ,final            
  ]
  {aipproc}

\layoutstyle{6x9}

\newcommand{\zaa}{Astron.~Astrophys.}

\newcommand{\zapj}{Astrophys.~J.}

\newcommand{\zapjs}{Astrophys.~J.~S.}

\newcommand{\zepj}{European Physical Journal}
\newcommand{\znp}{Nucl.~Phys.}
\newcommand{\zpl}{Phys.~Lett.}
\newcommand{\zpr}{Phys.~Rev.}
\newcommand{\zprl}{Phys.~Rev.~Lett.}

\newcommand{\zmnras}{Mon. Not. R. Astron. Soc.}

\newcommand{\gap}{\mathrel{ \rlap{\raise.5ex\hbox{$>$}}
                    {\lower.5ex\hbox{$\sim$}}  } }
\newcommand{\lap}{\mathrel{ \rlap{\raise.5ex\hbox{$<$}}
	            {\lower.5ex\hbox{$\sim$}}  } }

\begin{document}

\title{Nucleosynthesis in novae: experimental progress in the determination of nuclear reaction
rates}



\author{Alain Coc}{
  address={CSNSM, CNRS/IN2P3, Universit\'e Paris Sud,
B\^atiment 104, F--91405 Orsay Campus, France}
}



\begin{abstract}
The sources of nuclear uncertainties in nova nucleosynthesis have been identified using
hydrodynamical nova models.
Experimental efforts have followed and significantly reduced those uncertainties. 
This is important for the evaluation of nova contribution to galactic chemical evolution,
gamma--ray astronomy and possibly presolar grain studies. In particular, estimations of 
expected gamma--ray fluxes are essential for the planning of observations with existing or 
future satellites. 
\end{abstract}

\maketitle


\section{Introduction}

Novae are thermonuclear runaways occurring at the surface of a white dwarf accreting
hydrogen rich matter from its companion in a close binary
system\cite{Sta72,Geh98,JH98,JH07}. 
Material from the
white dwarf $^{12}$C and $^{16}$O (CO nova) or $^{16}$O, $^{20}$Ne plus some 
Na, Mg and Al isotopes (ONe nova) provide the seeds for the operation of the CNO cycle and
further nucleosynthesis. Novae are supposed to be at the origin of galactic $^{15}$N and 
$^{17}$O and
contribute to the galactic chemical evolution of $^7$Li and $^{13}$C. In addition they
produce radioactive isotopes that could be detected by their gamma--ray emission:
$^7$Be (478~keV), $^{18}$F ($\le$511~keV), $^{22}$Na (1.275~MeV) and $^{26}$Al (1.809~MeV). 
The yields of these isotopes depend strongly on the hydrodynamics of the explosion but
also on nuclear reaction rates involving stable and radioactive nuclei. 
Tests of sensitivity to the reaction rates uncertainties have been
done using parametrized\cite{Wor94}, semi--analytic\cite{Coc95}, 
post--processed\cite{Ili02} nova models but also with a 1-D hydrodynamical model. 
Indeed, in a series of papers the impact of nuclear uncertainties in the hot-pp
chain\cite{Be96}, the hot-CNO cycle\cite{F00}, the Na--Mg--Al region\cite{NaAl99} and
Si--Ar region\cite{SiAr01} have been investigated with the Barcelona (SHIVA) hydrocode.
In this way, the temperature and density profiles, their time evolution, and the effect 
of convection time scale were fully taken into account. 
The nuclear reaction rates whose uncertainties affected most nova
nucleosynthesis having been identified, many nuclear physics experiments were
conducted to reduce these uncertainties. In this review, we will shortly summarize the
experimental progress made in this domain.

\section{Hot CNO cycle}

The hot--CNO cycle deserves special attention as it is the main source of energy for
both type of novae and is the source for the production of $^{13}$C, $^{15}$N, 
$^{17}$O (galactic chemical evolution) and $^{18}$F (gamma--ray astronomy).
The positrons produced by $^{18}$F $\beta^+$ decay annihilate and are the dominant source of
gamma rays during the first hours of a nova explosion\cite{GG98}.  
Following a series of hydrodynamical calculations, major nuclear uncertainties on the
production of $^{17}$O and $^{18}$F were pointed out in Ref.~\cite{F00} (hereafter
CJHT).

\begin{figure}[h]
\includegraphics[height=.5\textheight]{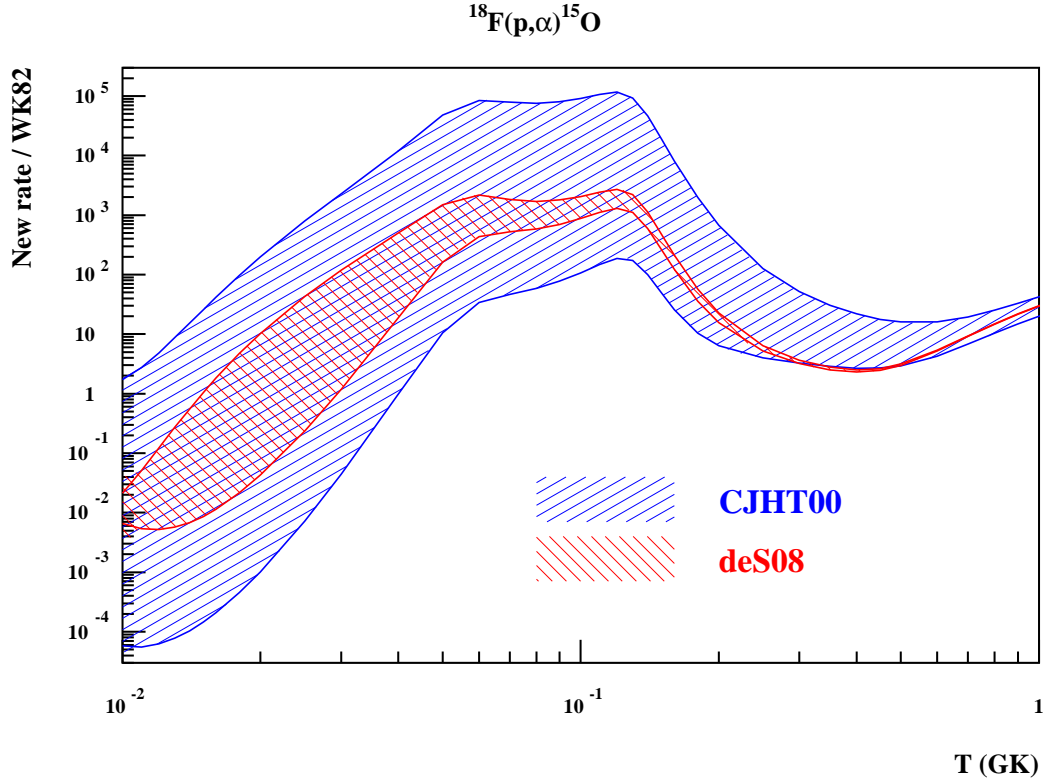}
\caption{Reduced relative uncertainty\protect\cite{f18pa} on the 
$^{18}$F(p,$\alpha)^{15}$O reaction rate. (An update of
Fig.~6 in CJHT\protect\cite{F00}.)}
\label{f:f18pa-sv}
\end{figure}

In particular, the $^{18}$F(p,$\alpha)^{15}$O reaction was recognized as the
main source of uncertainty for the $^{18}$F production, due to the unknown 
contributions of two low energy resonances. 
They are postulated at 8 and 38 keV following the spectroscopic work of
Utku et al.\cite{Utk98} who identified two $^{19}$Ne levels at $E_X$ = 6.419 and 
6.449 MeV and assumed that they are the analogs of the $E_X$ = 6.497 and 
6.528 MeV 3/2$^+$ $^{19}$F levels. 
Two (d,p) transfer reaction experiments, in inverse kinematics with a $^{18}$F beam, 
were conducted at Louvain--la--Neuve (LLN)\cite{dp} and Oak--Ridge (ORNL)\cite{Koz06}. 
They enabled to extract the neutron spectroscopic factor(s) of the (experimentally
unresolved) $^{19}$F analog levels. 
Since the one to one assignments of these analog levels separated by only 30 keV is not 
settled, even when assuming the equality of spectroscopic factors between analog levels, 
it is not possible yet to determine the two resonance relative contributions.
However, the extracted spectroscopic factor value imply that they must be included
in the calculation of the reaction rate.

In addition, if the spin and parity assignment is correct they should interfere with 
an other 3/2$^+$ broad resonance located at $E_R$ = 665 keV. 
This latter resonance and the 330 keV, 3/2$^-$
have been precisely measured directly\cite{Coz95,Bar02} at ORNL and LLN.
Interferences between the 8, 38 and 665 keV are expected to have a maximum effect
right in the energy range of interest for nova. It is hence extremely important to
determine their constructive or destructive nature. Until recently, the only
constraint in this energy region was provided by an {\em off--resonance} 
measurement\cite{Bar02} at 380 keV with a large error bar. A new direct measurement
of the $^{18}$F(p,$\alpha)^{15}$O cross section was recently performed at
LLN\cite{f18pa} at $E_{CM}$ = 726, 666, 485 and 400 keV. 
The higher energies correspond to the top of
the 665 keV resonances while the lower ones with respectively 180 and 35 counts are
located close to the limit of the interference region. With these new results,
R-matrix calculations including up to four 3/2$^+$ levels were performed to 
help constrain the S--factor. Even though more experiments are needed in the
interference region,when taking into account those recent 
measurements\cite{dp,Koz06,Bar02,f18pa} the reduction of the uncertainty on the 
$^{18}$F(p,$\alpha)^{15}$O reaction rate is important as shown in 
Figure~\ref{f:f18pa-sv}. In CJHT, the uncertainty on $^{18}$F production
due to this reaction was a factor of $\sim$300; it can now be estimated to a 
factor of $\sim$10.

\begin{figure}[h]
\includegraphics[height=.5\textheight]{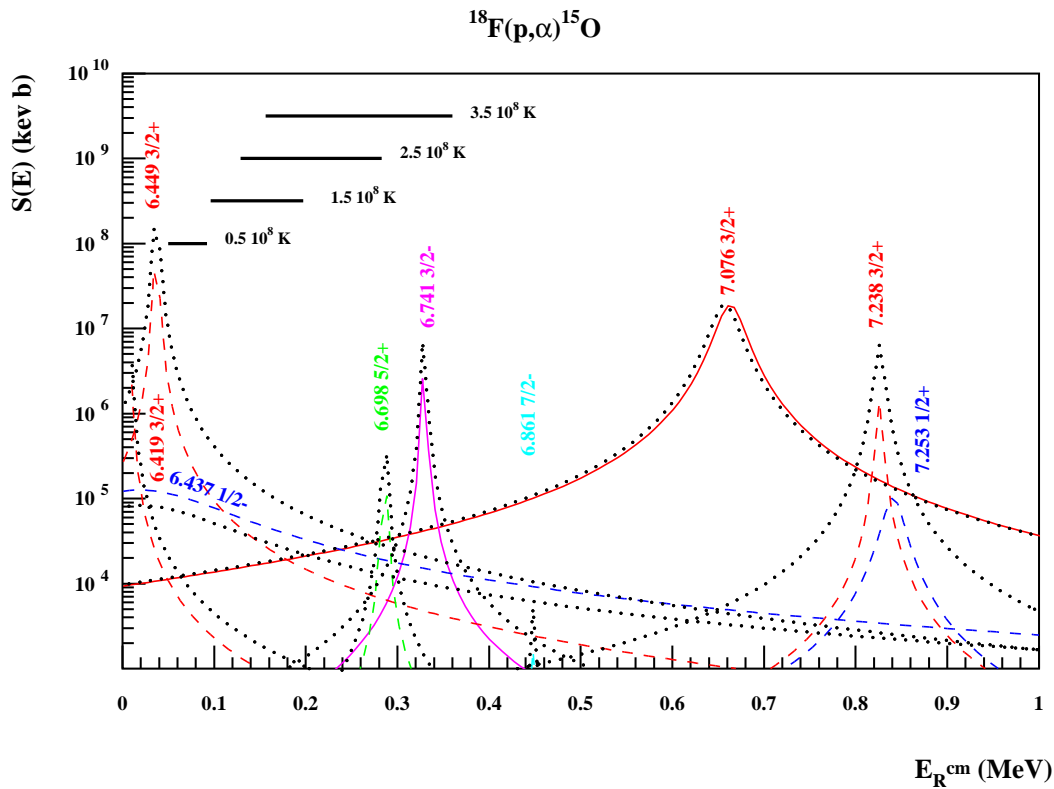}
\caption{(On--line only.) Contributions of known $^{19}$F levels to $^{18}$F(p,$\alpha)^{15}$O
astrophysical factor (color) compared to the previous situation (black dots)
from Fig.~4 in CJHT\protect\cite{F00}.}
\label{f:f18pa-se}
\end{figure}

We have up to now assumed that the reaction rate is dominated by the three 3/2$^+$ and
the 3/2$^-$ resonances but the comparison with $^{19}$F spectrum suggest that several
levels are missing in $^{19}$Ne. In particular two 1/2$^+$ ($\ell$=0) broad levels
have been
predicted by microscopic\cite{Duf07} calculations, one at $\approx$1 MeV above 
and another below threshold. If they exist they would lead to a significant
contribution in the relevant energy range. Data analysis of an inelastic scattering
experiment performed at LLN could provide information on this possible 
$\approx$1 MeV level\cite{fos}.

\begin{figure}[h]
\includegraphics[height=.5\textheight]{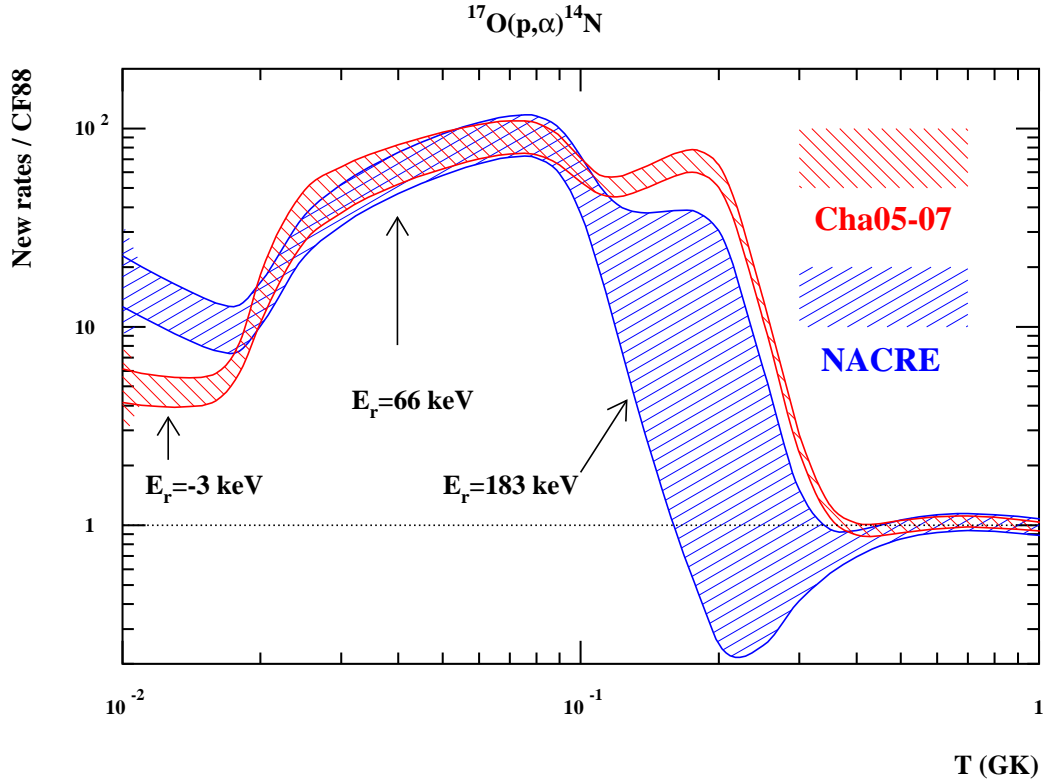}
\caption{Reduced relative uncertainty on the 
$^{17}$O(p,$\alpha)^{14}$N reaction rate\protect\cite{PAPAP}. (An update of
Fig.~8 in CJHT\protect\cite{F00}.)}
\label{f:o17pa}
\end{figure}

The $^{17}$O(p,$\alpha)^{14}$N and $^{17}$O(p,$\gamma)^{18}$F reaction were also identified
as sources of uncertainties for the production of $^{18}$F. The latter leads to the 
formation of $^{18}$F from the $^{16}$O seed nuclei trough the
$^{16}$O(p,$\gamma)^{17}$F($\beta^+)^{17}$O(p,$\gamma)^{18}$F chain while the former diverts
the flow reducing both $^{18}$F and $^{17}$O yields. According to the NACRE
compilation\cite{NACRE}, the uncertainty on these rates came from the, at that time unobserved,
resonance around 190~keV resulting in a factor of $\sim$10 additional uncertainty on  
$^{18}$F production\cite{F00}. The NACRE rates were based on experimental data which 
were found to be inaccurate after several measurements performed first at LENA\cite{LENA} and
in Orsay\cite{PAPAP}. There is now a good agreement, on the resonance energy
(e.g. 183.2$\pm$0.6~keV\cite{PAPAP}) and (p,$\alpha$) strength 
(e.g. 1.6$\pm$0.2~meV\cite{PAPAP}) but a small discrepancy concerning the (p,$\gamma$) 
strength (1.2$\pm$0.2~$\mu$eV\cite{LENA} and 2.2$\pm$0.4~$\mu$eV\cite{PAPAP}).

\begin{figure}[h]
\includegraphics[height=.5\textheight]{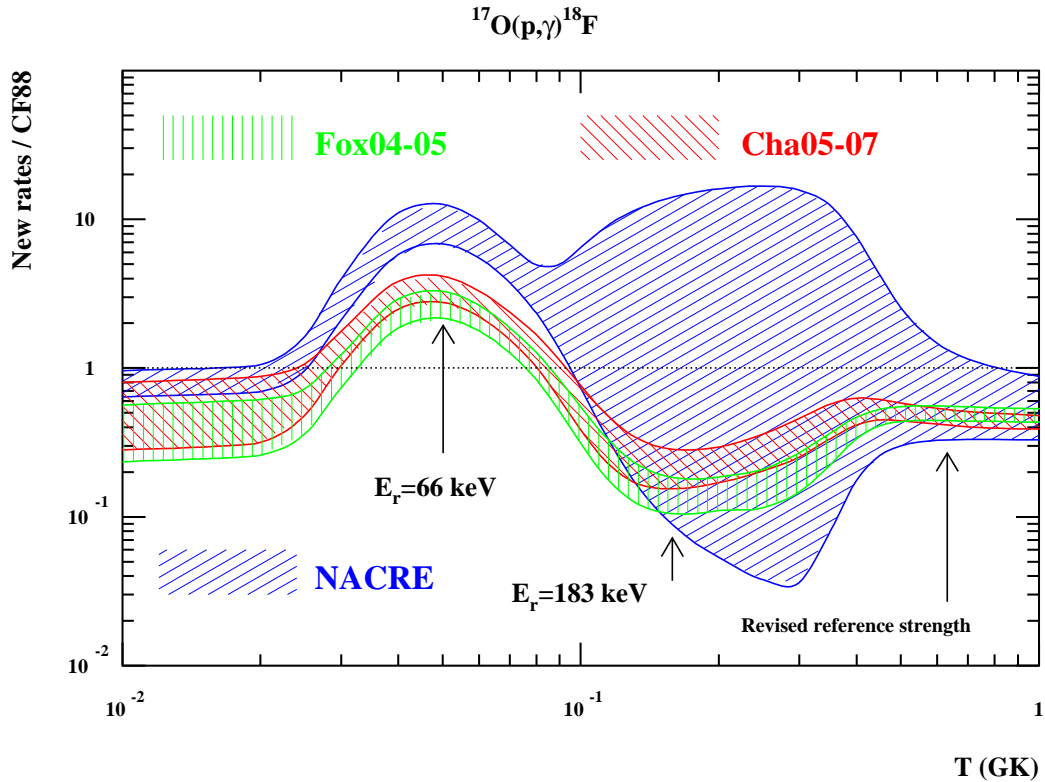}
\caption{Reduced relative uncertainty according to Chafa et al.\protect\cite{PAPAP} 
or Fox et al.\protect\cite{LENA}
on the $^{17}$O(p,$\gamma)^{18}$F reaction rate. 
(An update of Fig.~9 in CJHT\protect\cite{F00}.)}
\label{f:o17pg}
\end{figure}

Figures~\ref{f:o17pa} and ~\ref{f:o17pg} display the evolution of the $^{17}$O+p rates since
the NACRE compilation showing that they are now both known with sufficiently good accuracy for
nova applications. At nova temperatures, the $^{17}$O(p,$\gamma)^{18}$F rate is lower while 
the $^{17}$O(p,$\alpha)^{14}$N rate is higher. As a result, the $^{18}$F and $^{17}$O nova
production is smaller. (See Fox et al.\cite{LENA} for reanalysis of 
of the 66~keV and subthreshold resonance contributions.)

The reaction rates involved in $^{18}$F and $^{17}$O are now much better known and new
hydrodynamical calculations are underway to update their yields but also to better 
understand their nucleosynthesis.

\section{Other regions or reactions}
 
Nuclear uncertainties on the production of $^7$Li and $^7$Be are negligible compared with 
the hydrodynamics (rise time in temperature)\cite{Be96}.  
Even though some nuclear reaction rates are still uncertain, leaks from the CNO cycle
are negligible at novae temperatures. In particular, experimental data on
the $^{15}$O($\alpha,\gamma)^{19}$Ne\cite{Tan07} and 
$^{19}$Ne(p,$\gamma)^{20}$Ne\cite{Cou04} 
are now sufficient
to rule out any significant nuclear flow out the CNO cycle. Production of heavier
elements rely on the presence of $^{20-22}$Ne, $^{23}$Na,  $^{24-26}$Mg and $^{27}$Al
in ONe white dwarfs.
 
\subsection{$^{22}$Na production}

The decay of $^{22}$Na ($\tau_{1/2}$ = 2.6 y) is followed by the emission of a 
1.275 MeV photon. Observations have up to now only provided upper limits, compatible 
with model predictions, for this gamma ray emission. Its detection remains a goal for 
present (Integral) and future gamma-ray observatories. Calculating the expected 
$^{22}$Na yields used to be hampered by the nuclear uncertainties on the 
$^{21}$Na(p,$\gamma)^{22}$Mg and  $^{22}$Na(p,$\gamma)^{23}$Mg reaction 
rates\cite{NaAl99}. 

Destruction of $^{22}$Na in nova proceeds through the $^{22}$Na(p,$\gamma)^{23}$Mg 
reaction. Since the NACRE compilation a Gammasphere experiment\cite{Jen04} and a $\beta$--decay
experiment\cite{Iac06} have improved
the $^{23}$Mg spectroscopy, reducing by a factor of $\sim$10 the uncertainty below  
10$^8$~K. Unfortunately, they also discovered a new level which could lead to a yet
unobserved resonance at 190~keV. The rate uncertainty at nova temperatures remains
large: a factor of $\sim$10. 

\begin{figure}[h]
\includegraphics[height=.5\textheight]{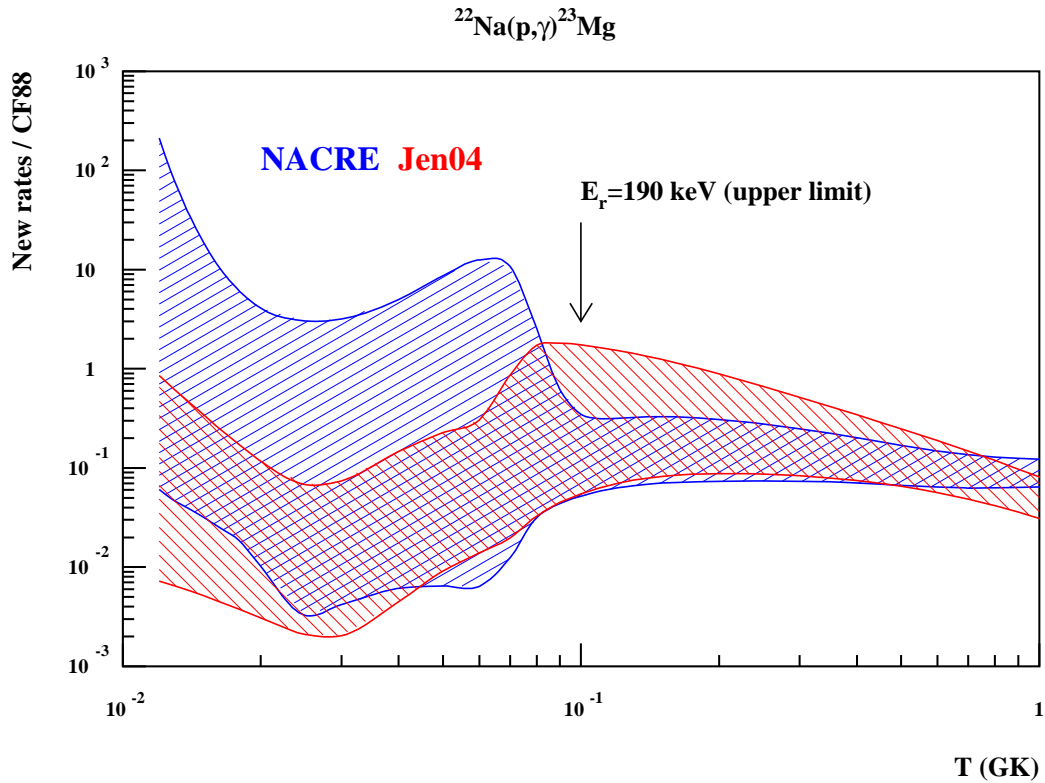}
\caption{(On--line only.) Evolution of the $^{22}$Na(p,$\gamma)^{23}$Mg reaction rate
uncertainty since Ref.~\cite{NaAl99}.}
\label{f:na22pg}
\end{figure}

Photodisintegration of $^{22}$Mg, important at nova temperatures, prevents
further processing but the $^{21}$Na(p,$\gamma)^{22}$Mg reaction remains important for  
$^{22}$Na production. It is a branching point between  
$^{21}$Na(p,$\gamma)^{22}$Mg($\beta^+)^{22}$Na and 
$^{21}$Na($\beta^+)^{21}$Ne(p,$\gamma)^{22}$Na, affecting the timescale and hence the
$^{22}$Na production\cite{NaAl99}. The uncertainty on this rate used to come from 
the unknown contributions of three unobserved resonances associated with
the $E_X$ = 5.714, 5.837 and 5.962~MeV $^{22}$Mg levels. Thanks to experiments 
conducted at the TRIUMF-ISAC facility with a $^{21}$Na beam, $i$) the $E_R$ = 206~keV 
($E_X$ = 5.714~MeV) resonance strength has been precisely measured\cite{Bis03} and 
$ii$) the contribution of the others was found to be negligible\cite{DAu04}.

\subsection{$^{26}$Al production}

With its long lifetime, $^{26g.s.}$Al ($\tau_{1/2}$ = 0.717 My) a single nova gamma 
ray emission (1.809~MeV) is far too faint to be observable but novae can contribute 
to the accumulation of this isotope in the Galaxy. 
The major nuclear uncertainties affecting its production were
identified to be the $^{25}$Al(p,$\gamma)^{26}$Si and  
$^{26g.s.}$Al(p,$\gamma)^{27}$Si\cite{NaAl99}.

The $^{26g.s.}$Al(p,$\gamma)^{27}$Si reaction governs $^{26}$Al ground state destruction
in novae. For nucleosynthesis calculations, its rate was often adopted from the 
{\em unpublished} work of Vogelaar. The NACRE compilation excluding unpublished results
assigns a large uncertainty to this rate at nova temperature. The strength of the
$E_R$ = 188~keV resonance was at the origin of this uncertainty and influenced 
directly the $^{26}$Al production \cite{NaAl99}. It has now been measured directly with a 
$^{26}$Al beam at the TRIUMF-ISAC facility and found\cite{Rui06} to be within a factor of 1.6 from the
unpublished Vogelaar's value. Nevertheless, the uncertainty remains large (orders of 
magnitudes) below 10$^8$~K because of the lack of information on 
lower energy resonances.

\begin{figure}[h]
\includegraphics[height=.5\textheight]{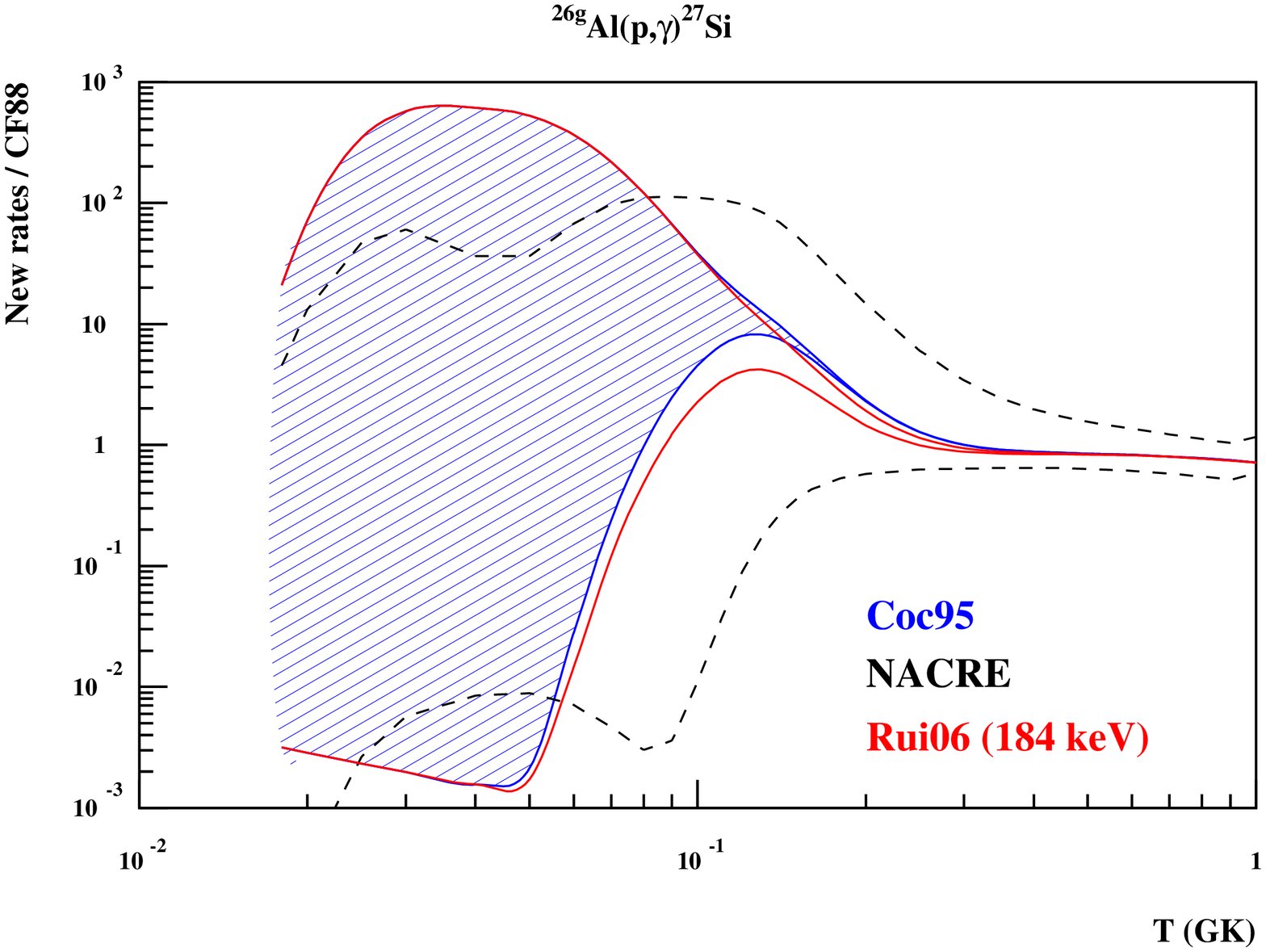}
\caption{(On--line only.) Evolution of the $^{26}$Ap(p,$\gamma)^{27}$Si reaction rate
uncertainty since Ref.~\cite{NaAl99}.}
\label{f:al26pg}
\end{figure}

Depending on the initial $^{24}$Mg abundance, the $^{25}$Al(p,$\gamma)^{26}$Si can
have a crucial role in the formation of $^{26g.s.}$Al as it provides a diversion from
the $^{24}$Mg(p,$\gamma)^{25}$Al($\beta^+)^{25}$Mg(p,$\gamma)^{26g.s.}$Al flow.
Following $^{25}$Al(p,$\gamma)$, $^{26}$Si can either decay to the short lived 
isomeric $^{26}$Al level or be destroyed by subsequent proton capture\cite{Togano}.
In either cases, it bypasses the long lived $^{26}$Al and gamma emitting ground 
state. (At nova temperature, $^{26}$Al isomer and ground states have to be
considered as separate species\cite{Thermal}.)
Orders of magnitudes uncertainties arose from missing $^{26}$Si levels,
in particular a 3$^+$ ($\ell$=0). 
Spectroscopic studies\cite{Bar02a,Cag02,Par04,Bar06,Sew07} of $^{26}$Si have lead to the
localization of a 4$^+$ and a 1$^+$ level and the probable localization of the  
3$^+$ level at $E_X$ = 5.912~MeV and 0$^+$ level at $E_X$ = 5.946~MeV.
The corresponding resonance strengths have not been measured, but the uncertainty 
on this rate has been considerably reduced.

\begin{figure}[htb]
\includegraphics[width=8.cm]{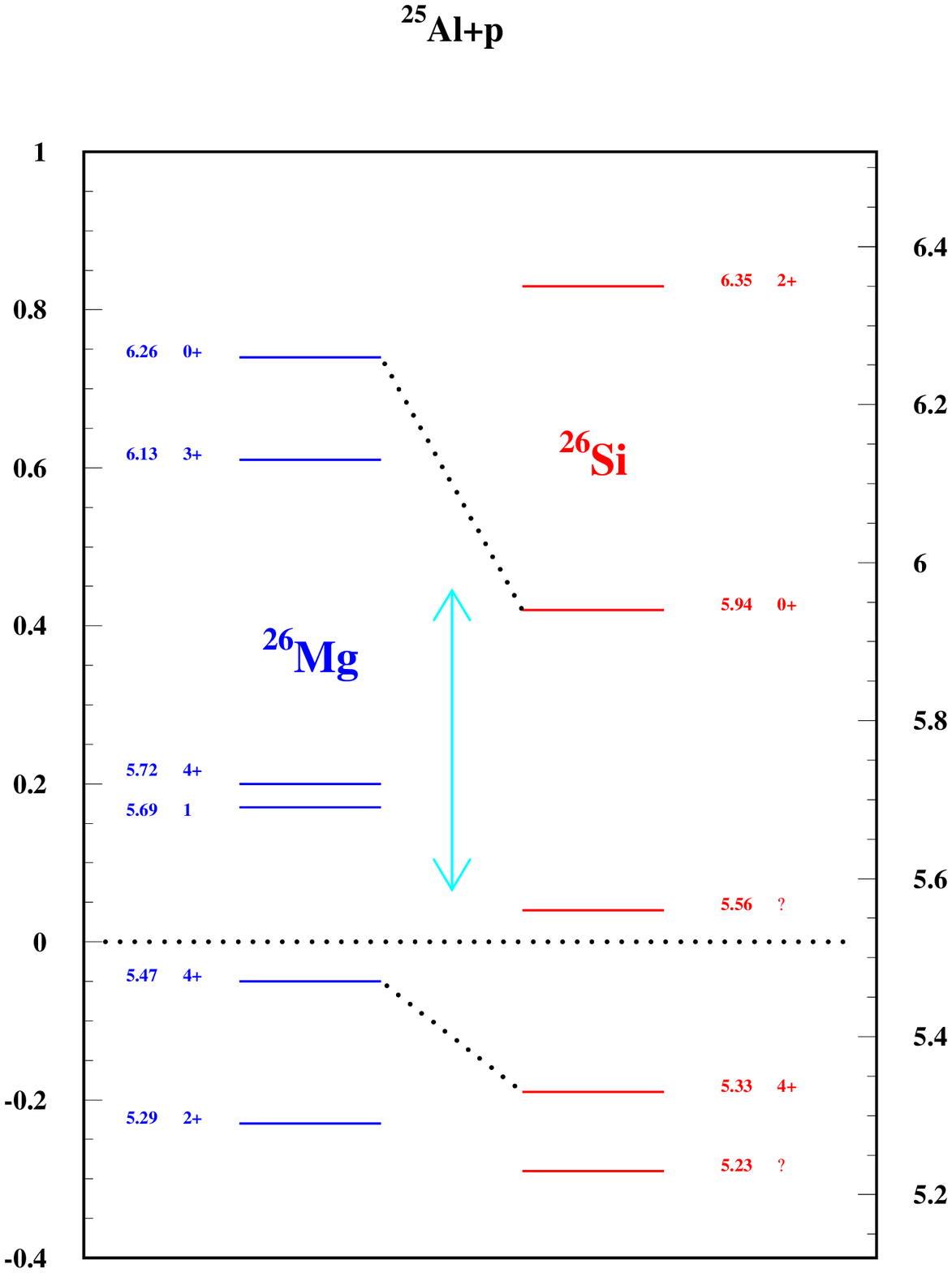}
\includegraphics[width=8.cm]{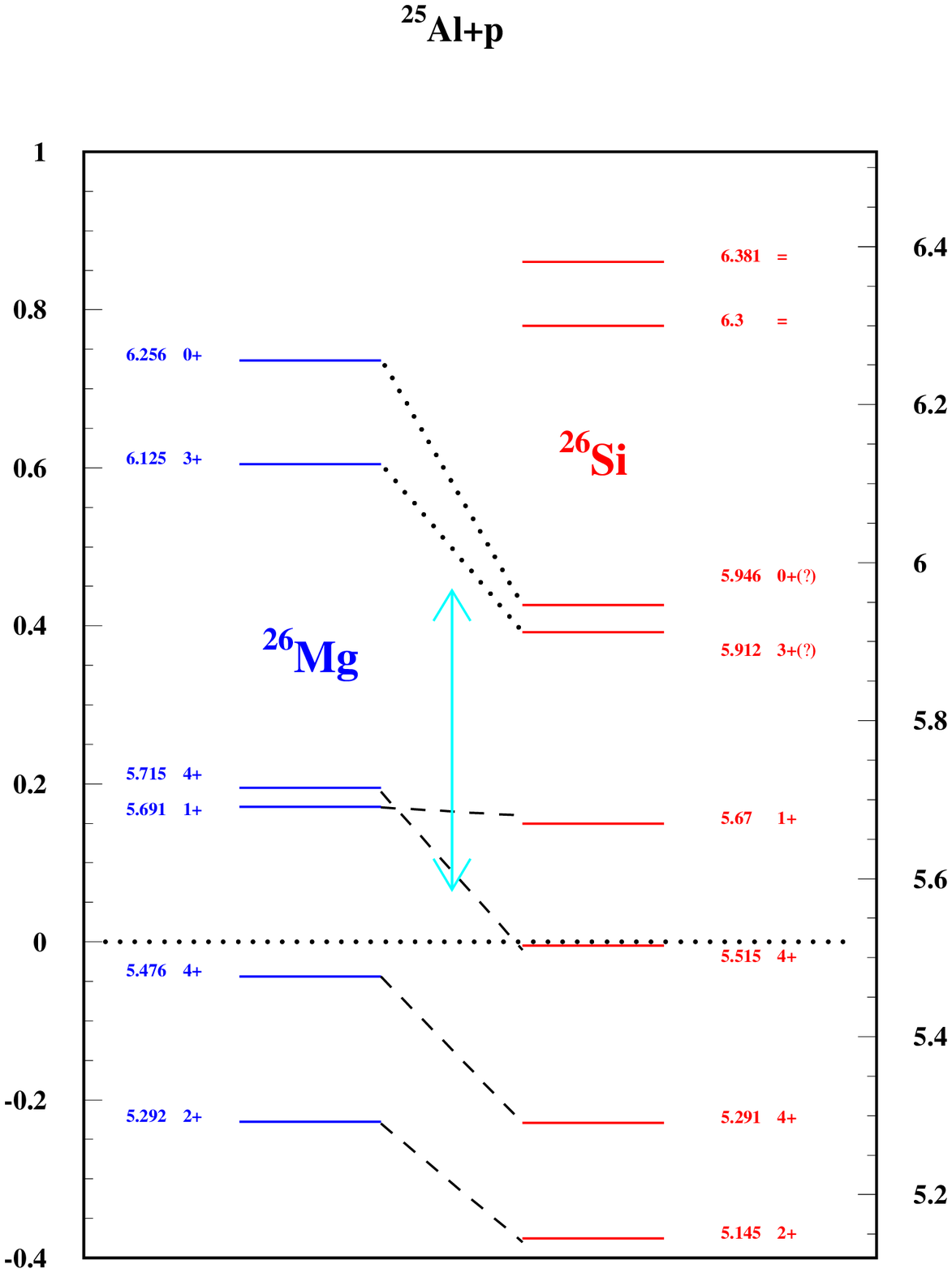}
\caption{(On--line only.) Progress in $^{26}$Si spectroscopy for the 
$^{25}$Al(p,$\gamma)^{26}$Si reaction rate since Ref.~\cite{NaAl99}. The arrow
represent the region of interest for nova nucleosynthesis.}
\label{f:al25pg}
\end{figure}

\subsection{Heavy elements production}

There is normally no significant amount of elements beyond aluminum found in white dwarfs
composition. Hence, the production of "heavy elements", i.e. from silicon to
argon, rely on the
nuclear flow out of the Mg-Al region through $^{28}$Si and subsequently 
through $^{30}$P whose
relatively long lifetime ($\tau_{1/2}$= 2.5~mn) may halt the flow unless the 
$^{30}$P(p,$\gamma)^{31}$S is
fast enough. This reaction is also important to calculate the silicon isotopic ratios
to be compared to values measured in some presolar grains that may have a nova 
origin\cite{Ama01}. Due to the limited spectroscopic data available for the $^{31}$S
nucleus, up to now an Hauser--Feshbach rate was used in nova nucleosynthesis calculations.
This statistical model, assuming a high level density, is certainly
not appropriate for such a low
mass nucleus and temperature domain. The uncertainty was difficult to determine but
two orders of magnitudes was the usual estimate. The $^{31}$S spectroscopy is not yet
completed in the range of excitation energy important for nova but about ten levels have
been observed\cite{Jen05,Jen06,Kan06,Ma07,Wre07} in that region with spins and parities 
generally assigned. The relatively high level density prevented all these levels to be
experimentally resolved so that confirmations would be welcomed. The resonance strength
used to calculate the thermonuclear reaction rate are also obtained by 
{\em assuming} typical values for
spectroscopic factors. The resulting reaction rate, even though still uncertain, present
a significant improvement and seems\cite{Wre07} close to the Hauser--Feshbach one used
in previous nucleosynthesis calculations.   

\begin{figure}[htb]
\includegraphics[width=8.cm]{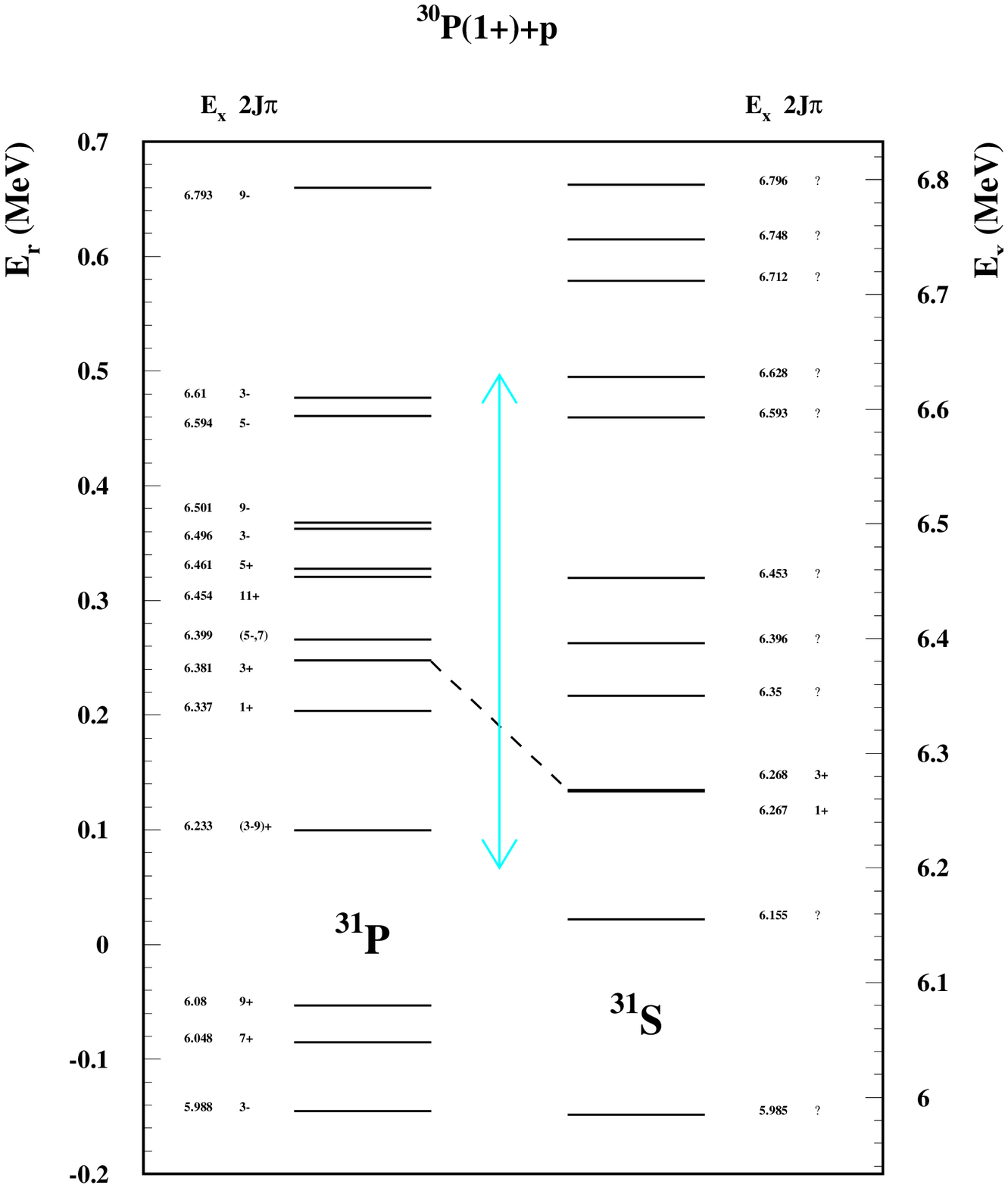}
\includegraphics[width=8.cm]{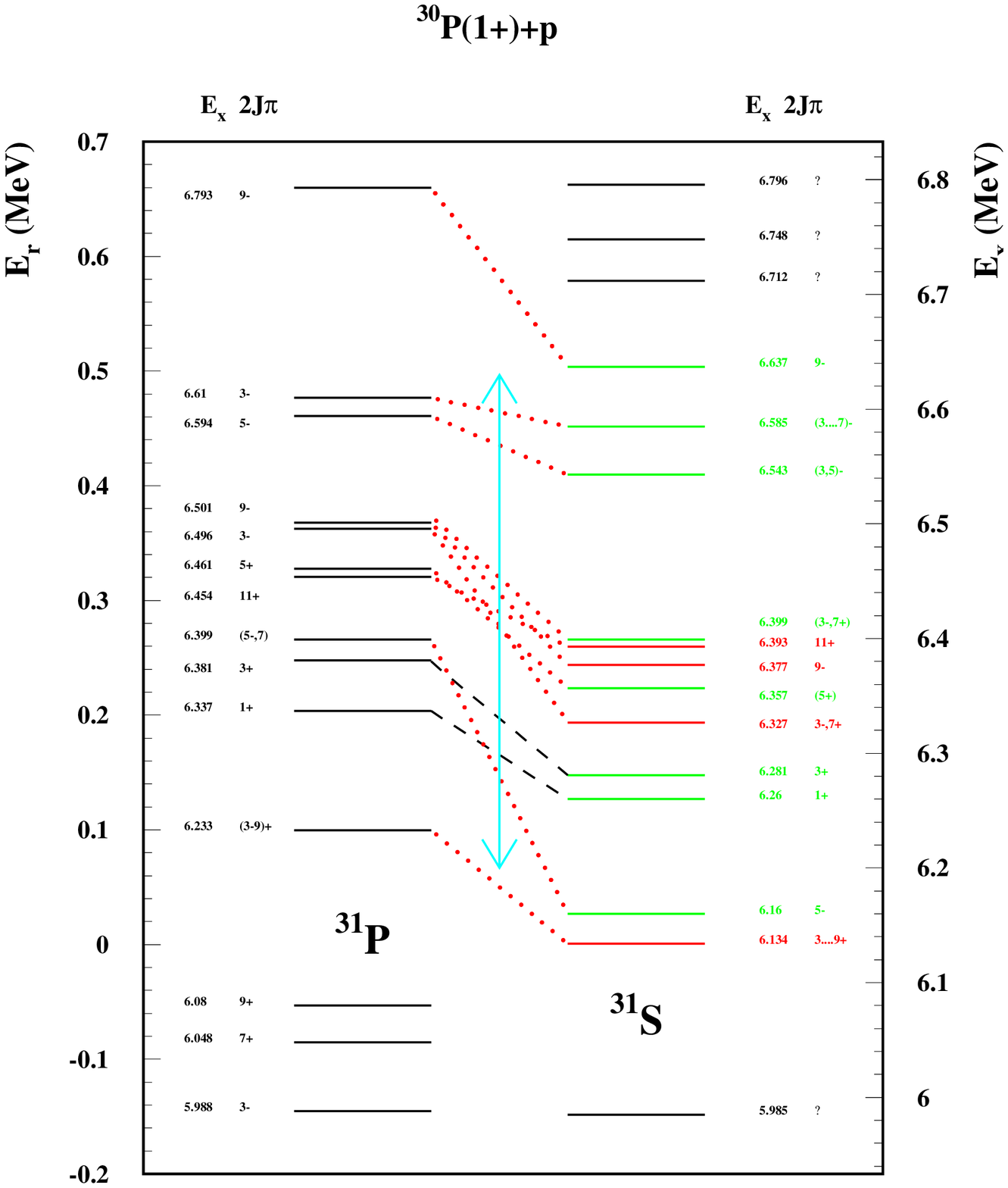}
\caption{(On--line only.) Progress in $^{31}$S spectroscopy for the 
$^{30}$P(p,$\gamma)^{31}$S reaction rate since Ref.~\cite{SiAr01}.}
\label{f:p30pg}
\end{figure}

\section{Conclusions}

Detailed calculations performed with the SHIVA hydrodinamical code
have enable the identification of nuclear uncertainties affecting 
nova nucleosynthesis. We see that less than ten years after, 
great progress have been made thanks to experimental efforts, in particular for the
$^{17}$O(p,$\gamma)^{18}$F, $^{17}$O(p,$\alpha)^{14}$N, 
$^{18}$F(p,$\gamma)^{19}$Ne, $^{18}$F(p,$\alpha)^{15}$O,
$^{21}$Na(p,$\gamma)^{22}$Mg, $^{22}$Na(p,$\gamma)^{23}$Mg 
$^{25}$Al(p,$\gamma)^{26}$Si, $^{26g.s.}$Al(p,$\gamma)^{27}$Si
and $^{30}$P(p,$\gamma)^{31}$S reactions that were identified as the most influential.
However, further efforts are required for the 
$^{22}$Na(p,$\gamma)^{23}$Mg, $^{25}$Al(p,$\gamma)^{26}$Si, $^{30}$P(p,$\gamma)^{31}$S 
reactions and especially for the $^{18}$F(p,$\alpha)^{15}$O reaction.
For this last reaction where contribution of interfering broad resonance tails are  
essential, progress should come from direct measurements with intense $^{18}$F beam
(TRIUMF) or from indirect (THM\cite{THM}) measurement planned at the CRIB of the
Center for Nuclear Studies (Wako).


\begin{theacknowledgments}
I am indebted to Margarita Hernanz and Jordi Jos\'e for a now more than twelve years 
collaboration on nova nucleosynthesis and to Nicolas~de~S\'er\'eville for 
frequent discussions.
Many thanks also to Carmen Angulo, Christian Iliadis, 
Fa\"{\i}rouz Hammache, Fran\c{c}ois de Oliveira Santos and Claudio Spitaleri 
for long time collaborations. 
\end{theacknowledgments}

\bibliographystyle{aipproc}   


\end{document}